# Mid-infrared time-resolved photoconduction in black phosphorus

Ryan J. Suess[1,2,a)], Edward Leong[2], Joseph L. Garrett[1,3], Tong Zhou[4], Reza Salem[4], Jeremy N. Munday[1,2], Thomas E. Murphy[1,2], and Martin Mittendorff[1]

[1]*Institute for Research in Electronics & Applied Physics, University of Maryland, College Park, Maryland, 20742, USA*

[2]*Department of Electrical & Computer Engineering, University of Maryland, College Park, Maryland, 20742, USA*

[3]*Department of Physics, University of Maryland, College Park, Maryland, 20742, USA*

[4]*Thorlabs, 10335 Guilford Rd, Jessup, Maryland, 20794, USA*

a) ryan.suess@gmail.com

**Abstract**

Black phosphorus has attracted interest as a material for use in optoelectronic devices due to many favorable properties such as a high carrier mobility, field-effect, and a direct bandgap that can range from 0.3 eV in its bulk crystalline form to 2 eV for a single atomic layer. The low bandgap energy for bulk black phosphorus allows for direct transition photoabsorption that enables detection of light out to mid-infrared frequencies. In this work we characterize the room temperature optical response of a black phosphorus photoconductive detector at wavelengths ranging from 1.56 μm to 3.75 μm. Pulsed autocorrelation measurements in the near-infrared regime reveal a strong, sub-linear photocurrent nonlinearity with a response time of 1 ns, indicating that gigahertz electrical bandwidth is feasible. Time resolved photoconduction measurements covering near- and mid-infrared frequencies show a fast 65 ps rise time, followed by a carrier relaxation with a time scale that matches the intrinsic limit determined by autocorrelation. The sublinear photoresponse is shown to be caused by a reduction in the carrier relaxation time as more energy is absorbed in the black phosphorus flake and is well described by a carrier recombination model that is nonlinear with excess carrier density. The device exhibits a measured noise-equivalent power of 530 pW·Hz$^{-1/2}$ which is the expected value for Johnson noise limited performance. The fast and sensitive room temperature photoresponse demonstrates that black phosphorus is a promising new material for mid-infrared optoelectronics.

# Introduction

Black phosphorous, a two-dimensional allotrope of phosphorus, has attracted interest because of its adjustable bandgap and favorable optical and electrical properties. Similar to graphene and other van der Waals materials, black phosphorus can be exfoliated into thin flakes from the bulk crystal[1]. Depending on the thickness of the flake, the band gap varies from about 2 eV for a single layer to 0.3 eV for more than 5 layers[2], with each layer consisting of a puckered hexagonal arrangement of phosphorus atoms. Large mobility values of up to 10,000 $cm^2/V\cdot s$[3] and a field effect that allows tuning of the Fermi level via a back gate, has enabled field-effect transistors with on-off ratios of several orders of magnitude and high speed, multispectral imaging sensors, and chemical detectors[4,5,6,7]. The field effect, high mobility, and low band gap make black phosphorus particularly attractive for optoelectronic devices in the near and mid-infrared (MIR) frequency range [6,8,9,10,11,12]. In the present work, the properties and potential of black phosphorus for mid-infrared photoconductive detection are explored. Photoconductive operation is verified via Kelvin probe force microscopy (KPFM), electrical transport measurements, and photocurrent measurements. Near-infrared autocorrelation measurements are used to quantify the intrinsic response time of the photoconduction. Subsequent time resolved measurements of the impulse response demonstrate nanosecond-scale detection at wavelengths up to 3.75 μm with a time response that is limited by the carrier lifetime. The noise performance is characterized and shows a thermally limited noise-equivalent power of 530 $pW\cdot Hz^{-1/2}$. Finally we compare the detector performance with existing MIR detector technologies.

# Device Fabrication and Characterization

Thin flakes of black phosphorus are produced via mechanical exfoliation onto a $SiO_2$ (300 nm) on Si substrate. We used a p-doped substrate with a resistivity of 250 Ω·cm, which allows for electrical back-gating of the device without causing significant absorption of the mid-IR wavelengths. To minimize exposure of the black phosphorus flake to ambient humidity, photoresist is spun immediately following the exfoliation. Standard photolithography with subsequent metal deposition (100 nm Au on top of 10 nm Cr) is employed to pattern electrical contacts to each flake, typically forming a 5 μm×5 μm black phosphorus channel. The crystallographic orientation of each flake was not measured or controlled during fabrication. After the lift off, the device is covered with a protective capping layer of 100 nm $Al_2O_3$ by atomic layer deposition. Chips used in the autocorrelation and impulse response measurements are wired to a SMA connector for high speed operation[13].

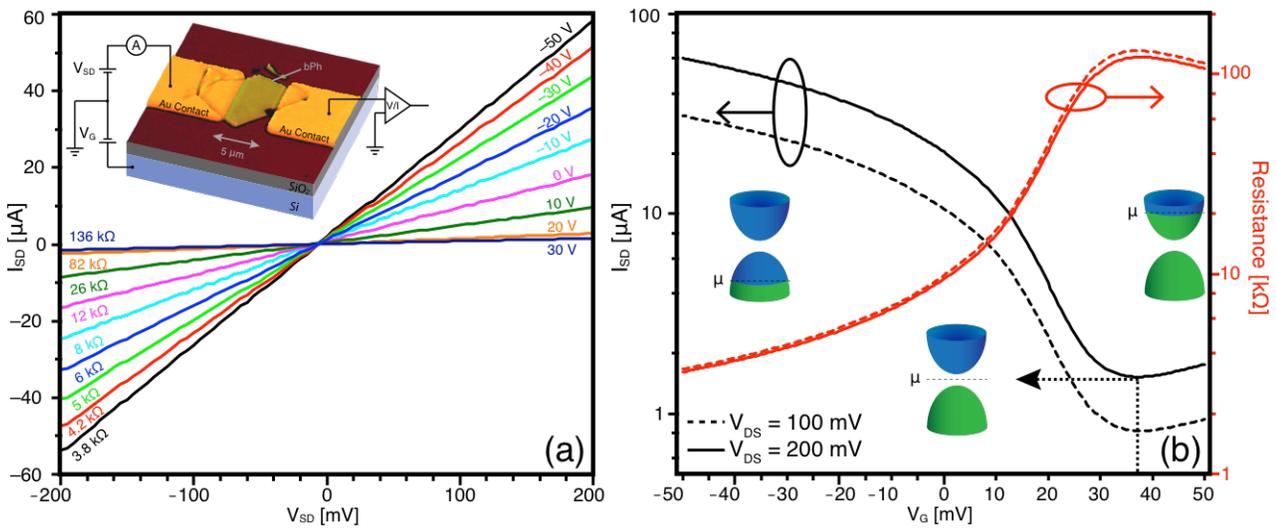

Figure 1. Unilluminated electrical characterization measurements. (a) Current-voltage characteristic curves for a typical black-phosphorus FET for various gate voltages (representative device optical micrograph with connection diagram shown inset). (b) Electrical transport measurements showing source-drain current (black) and resistance (red) as a function of gate voltage for two different bias voltages (dashed, solid lines).

The source-drain IV characteristic for a typical device is plotted in Fig. 1(a) for a range of gate voltages. The linear relationship between the applied bias and current indicates that the black phosphorus channel is resistive and does not exhibit any deviations that would suggest non-Ohmic contact between the black phosphorus and metal. The electrical transport through the channel is plotted in Fig. 1(b) along with schematic diagrams showing the chemical potential, μ, relative to the valance and conduction bands and band gap. From the gating curve we can estimate the field-effect charge carrier mobility[4]. At gate voltages below the charge neutral point ($V_{CNP}$=37 V) transport is dominated by holes that exhibit a mobility of 350 cm$^2$/V·s, while for gate voltages above $V_{CNP}$, the charge transport is dominated by electrons with an estimated mobility of 10 cm$^2$/V·s. Owing to the anisotropic in-plane conductivity of black phosphorus[3] and the uncontrolled orientation of each flake with respect to the contacts, a range of carrier mobilities were observed across the black phosphorus devices, typically 350-850 cm$^2$/V·s for holes and 10-100 cm$^2$/V·s for electrons.

Further insight into the electrical behavior of the device is gained by mapping the surface potential of the FET using heterodyne Kelvin probe force microscopy (H-KPFM) [14,15,16]. The surface potential provides information about the possible formation of interfacial electrostatic junctions, the existence of charge impurities, adsorbates and defects in the conduction channel, as well as verifies the potential gradient in the channel resulting from an applied bias, a condition required for photoconductive operation[17,18,19]. A plot of the topography, also obtained during the surface potential measurement, is shown in Fig. 2(a), while profiles of the channel averaged surface potential relative to the drain potential for different bias voltages are displayed in Fig. 2(b). The applied fields are separated from adsorbates and work function contrast by subtracting a null KPFM measurement, in which all the channels are grounded, from the data. As anticipated from the IV characteristic (cf., Fig. 1(a)), the surface potential profiles are linear across the channel for positive and negative source-drain bias, with no disruptions near the black phosphorus-metal interface, further confirming the Ohmic nature of the contacts[20]. The absence of surface potential perturbations near the contact region precludes the existence of a photovoltaic contribution to the photoresponse, for example, by asymmetric Schottky barriers formed at the gold-phosphorous interfaces. The surface potential fluctuations in the black phosphorus channel were observed to be 25 mV, much smaller than the typical applied bias potential of 200 mV, and equal to the minimum detectable voltage[16]. The instrument-limited potential fluctuations observed in our measurements show that there are no surface charge impurities in the black phosphorus channel within the accuracy of the measurement.

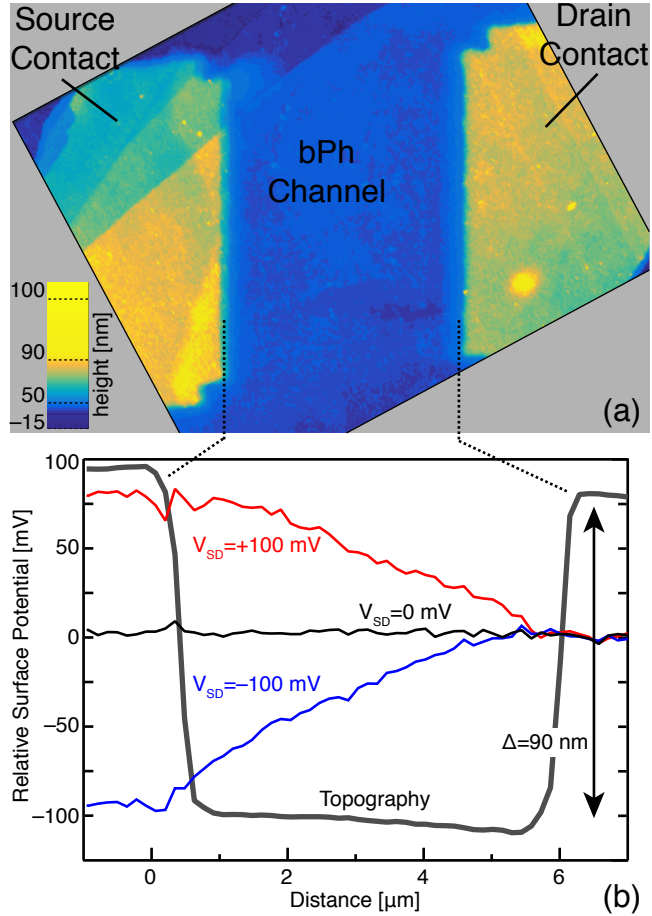

Figure 2. (a) Topography of a black phosphorus (bPh) channel for a typical device acquired during Kelvin probe force microscopy measurements. (b) Channel surface potential profiles shown under various bias conditions overlaid with topography profile (grey). A device with reduced cover layer oxide thickness of 20 nm is used for this measurement.

**Optical Photoresponse**

Several processes have been proposed to explain the photoresponse of black phosphorus detectors, including photovoltaic, photothermal, bolometric, and photoconductive mechanisms. As explained above, spatial measurements of the work function using heterodyne KPFM, together with electrical measurements show no evidence of a Schottky barrier or depletion region, indicating negligible contribution from photovoltaic processes. To assess the possibility of photothermal detection, we removed the DC bias and selectively illuminated one contact, which produces a photothermal current when hot, photoexcited electrons diffuse into the cooler contact. The photothermal signal, while present, was observed to be 2 orders of magnitude smaller than the photoconductive signal produced under DC bias. The photothermal contribution is further suppressed by employing global illumination. To explore possible bolometric contributions to the detection mechanism, we measured the electrical resistance at elevated temperatures, and observed that in the in the hole doped regime, where our device is operated, the resistance increases with substrate temperature (i.e., dR/dT>0) as a result of increased carrier scattering. This thermally-induced change in resistance is opposite to what we observe under optical illumination, indicating the photoconductive effect is the primary process governing the channel conductivity.

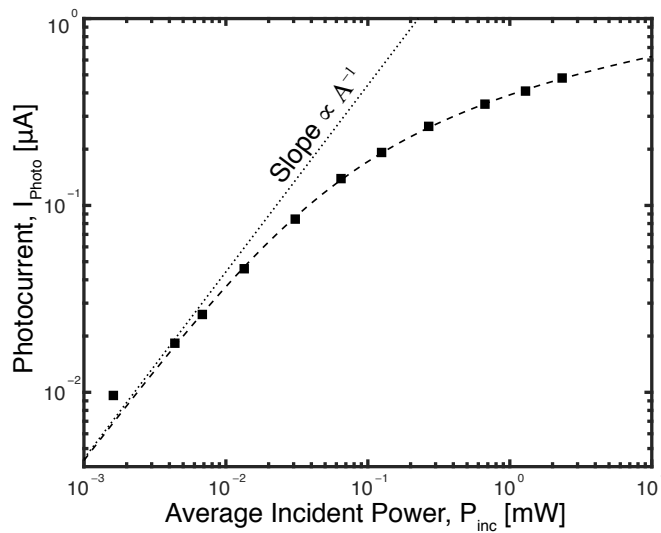

Figure 3. Near-infrared photoresponse of a black phosphorus detector showing the source-drain photocurrent as a function of average incident optical power. A back-gate voltage of 0 V and a source-drain bias voltage of 50 mV is used for this measurement. The dashed curve shows a fit to a carrier recombination model and its limit for low incident powers (dotted line).

The average current produced by a photoinduced change in the black phosphorus conductivity is characterized using a chopped, pulsed laser source with 1.56 µm wavelength, 100 fs pulses, and 100 MHz repetition rate focused to a spot size of 4 µm centered on the device. The signal resulting from the conductivity change is amplified using a transimpedance amplifier and measured with a lock-in amplifier. The detector is operated without a back-gate voltage and uses a source-drain bias of 50 mV.

The photocurrent shown in Fig. 3 exhibits a sublinear power dependence that can be explained by a photoexcited carrier lifetime that decreases with carrier concentration. The excess carrier density generated by the incident pulsed illumination increases linearly with power through direct transition photoabsorption. The lifetime of the photogenerated carrier population, however, is nonlinear with incident power as the recombination processes can have a quadratic or cubic dependence on the excess carrier density for radiative transitions or Auger processes, respectively. A carrier rate equation[21] is used to model the excess carrier concentration in the black phosphorus channel:

$$\frac{dn(t)}{dt} = G(t) - An(t) - Bn^2(t). \tag{1}$$

Here n(t) is the excess carrier density, G(t) is the carrier generation term, and A and B are the minority carrier lifetime and radiative recombination coefficients. Though higher order effects such as Auger recombination or spatial effects like diffusion can also play a role, we estimate these effects are small compared to the minority carrier lifetime and radiative components[22] which produce excellent agreement with the data using only two parameters (cf., Fig. 3). When the detector is illuminated with femtosecond optical pulses (as here), we may accurately model the carrier generation term G(t) in Eqn. 1 by an impulse, i.e,. G(t) = $n_0$ δ(t), where $n_0$ represents the initial, optically induced excess carrier concentration, which we take to be linearly proportional to the incident optical power. The measured current is proportional to the total charge that flows during the conductivity transient. The carrier concentration found from the rate equation is integrated in time to produce an average net photocurrent:

$$I_{Photo} = I_0 \ln\left(1 + \frac{n_0(P_{inc})}{n_{sat}}\right). \quad (2)$$

In Eqn. 2, $n_0(P_{inc})$ is the initial carrier concentration in the flake and is directly proportional to the incident power, $P_{inc}$, by the number of absorbed photons divided by the volume of the flake (estimated initial carrier concentrations are between $3\times10^{17}$ cm$^{-3}$ and $5\times10^{20}$ cm$^{-3}$ for the data shown in Fig. 3). The saturation density, $n_{sat}$, denotes the excess carrier concentration at which the photocurrent transitions from linear to sublinear power dependence and is given by, $n_{sat}$=A/B=$5.06\times10^{18}$ cm$^{-3}$. The term $I_0$=105 nA is a scaling term that depends on experimental parameters like source-drain bias and converts the integrated carrier density to a current. By fitting A from time resolved data (cf., Fig. 5(a)) we determine A=0.96 ns$^{-1}$ and B=$1.90\times10^{-10}$ s$^{-1}$·cm$^3$ (the latter is determined from knowing A/B as found from the fit in Fig. 3). The radiative recombination coefficient is in agreement with the value of $(2.3\pm0.3)\times10^{-10}$ s$^{-1}$·cm$^3$ obtained by time-resolved microwave conductivity measurements[22] and is comparable in magnitude to other low bandgap materials like InAs [23].

**Autocorrelation Measurements**

To measure the intrinsic speed of the black phosphorus photoconductive detector, we performed autocorrelation measurements using the photocurrent generated by ultrafast optical pulses. In this measurement, two identical laser pulses separated in time by a variable delay illuminate the black phosphorus channel while the average photocurrent is monitored. Two fiber lasers produce 100 fs pulses at a wavelength of 1.56 μm and have a synchronized repetition rate of 100 MHz. The lasers have identical power, but the repetition rate of one lasers is detuned from 100 MHz, resulting in a scanning delay up to 5 ns between the pulses emitted by each laser. As in the single-beam measurements presented in Fig. 3, the lasers were optically chopped and focused onto the black phosphorus channel while the resulting photoconductive signal was amplified by a transimpedance amplifier and measured using a lock-in amplifier. The photocurrent is plotted in Fig. 4(a) as a function of the time delay between the two pulses for a variety of linear polarization angles with a bias voltage of 200 mV and back gate voltage of 0 V. When the pulses have a maximum delay of ±5 ns, the photocurrent for an incident energy density of 300 μJ·cm$^{-2}$ exhibits values ranging from 0.9 μA to 2.2 μA, depending on the angle of linear polarization. The variation in photocurrent is the result of anisotropic optical absorption in black phosphorus[3], with the strongest absorption occurring when the angle of linear polarization is parallel to the high-mobility axis. The photocurrent at 0 ns delay is smaller than that observed at large delays for all polarization angles which is a consequence of the sublinear power dependence (cf., Fig. 3).

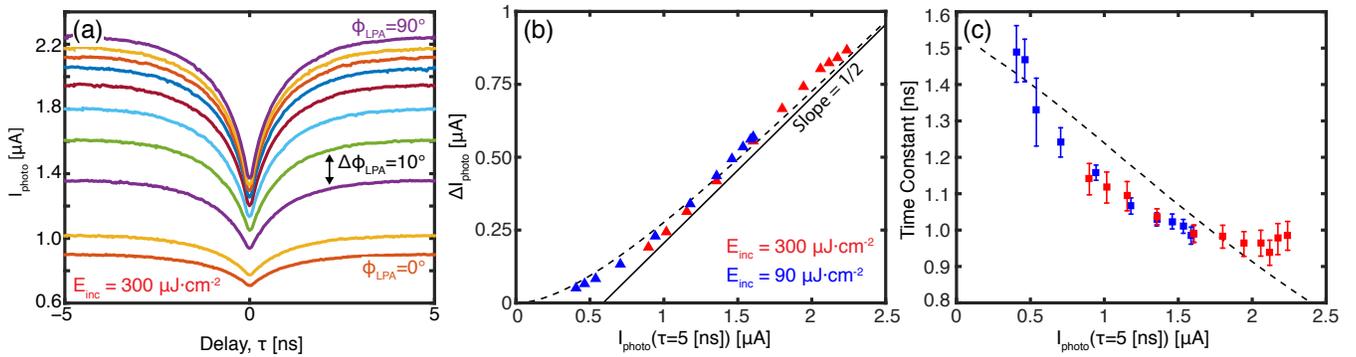

Figure 4. (a) Autocorrelation data showing photocurrent as a function of pulse delay for an incident energy density of 300 μJ·cm$^{-2}$. Solid curves are separated by 10° in linear polarization angle. (b) Difference in photocurrent at 0 ns and 5 ns delay plotted as a function of the photocurrent at a delay of 5 ns for incident energy densities of 90 µJ·cm$^{-2}$ (blue symbols) and 300 µJ·cm$^{-2}$ (red symbols). (c) Extracted time constant plotted against 5 ns delay photocurrent.

At delays much larger than the device response time, the two optical pulses from each laser contribute independently to produce a photocurrent that is the sum of what each laser would produce separately. When the pulses coincide in time (i.e., at $\tau$=0 ns), the cumulative photocurrent is smaller than when they are delayed, as shown in Fig. 4(a). The degree of photocurrent nonlinearity can be quantified by considering the relative change in photocurrent between the two cases of 0 ns and 5 ns delay.

Figure 4(b) shows the difference in photocurrent at 0 ns and 5 ns delay as a function of the background photocurrent (measured at a delay of 5 ns). Autocorrelation data taken at a lower incident energy density (blue symbols indicate 90 µJ·cm$^{-2}$) are also plotted. The difference in photocurrent increases with the background photocurrent which is a monotonically increasing function of absorbed power (cf., Eqn. 2). As predicted by the carrier recombination model, the data asymptotically approach a line with slope one-half, indicating the autocorrelation measurement cannot have a change in photocurrent more than half of the background value. The consistency of the two data sets shown in the figure indicates that the photocurrent difference appears to depend on the linear polarization angle only through the amount of power absorbed by the flake (i.e., the overlapping data points from each data set in Fig. 4(b) correspond to different polarization angles and incident optical powers). The dashed line plots the photocurrent expected from the excess carrier model using the same parameters determined by the fit shown in Fig. 3.

The time constant is obtained by fitting the autocorrelation measurement to a two-sided exponential and allows for measurement of the intrinsic timescale of the material nonlinearity without the influence of circuit parasitics[24]. Furthermore, the sublinear power dependence observed in the photocurrent can be understood by considering the time constant extracted from the autocorrelation data. The time constant shown in Fig. 4(c) is seen to decrease as a function of the background photocurrent. The decreasing time constant indicates that less total charge is integrated at higher absorbed energies and will therefore produce a smaller average photocurrent when both pulses are coincident. The time constant for this effect reaches a minimum value of approximately 950 ps for large values of the absorbed power. The time constant found from numerically simulating the autocorrelation experiment with identical parameters to those used previously is shown as a dashed line in the figure. The shortening time constant is also consistent with prior experimental results from purely optical pump-probe measurements that showed the emergence of a fast recombination peak as the optical fluence was increased[25]. We additionally measured the source-drain bias dependence of the carrier recombination rate. A 10% reduction was observed in the recovery time with increasing bias voltages up to 200 mV and is the result of some of the photoexcited carriers being swept out of the black phosphorus channel by the applied field. Devices using thinner flakes that are still within the bulk bandgap limit and oriented with the high mobility axis of the flake to be along the direction of the source-drain current, would allow for higher fields to be applied (if gated to minimum conductivity) and would result in higher carrier drift velocities. Decreasing the carrier transit time below the recombination lifetime would potentially allow for an additional reduction of the device recovery time[26].

**Infrared Impulse Response Characterization**

To directly measure the functional bandwidth of our device, the impulse response is measured at near- and mid-infrared frequencies at room temperature (T=300 K). In these measurements pulsed, infrared sources having wavelengths from the near- to mid-infrared illuminate the detector. All impulse response measurements are carried out with zero applied back gate with the device operating in the hole-doped regime. The detector is connected via a bias tee with a source-drain voltage of 200 mV, and the fast photoinduced change in resistance is recorded using a sampling oscilloscope with a bandwidth of 40 GHz.

Two different laser systems were used to characterize the near- and mid-infrared response of our device. In the near-infrared, we used the same femtosecond erbium-doped fiber laser that was used for the autocorrelation measurements presented above. The pulses were focused onto the sample using a high numerical aperture aspheric lens. For the mid-infrared measurements, a supercontinuum (SC) system consisting of a femtosecond pump laser and a dispersion-engineered indium fluoride fiber for SC generation was used. The pump source (Thorlabs FSL1950) provides 100 fs long pulses at 2.14 µm center wavelength, 50 MHz repetition rate, and 580 mW average power. The pump output is coupled into the SC-generating fiber that produces light with wavelengths spanning from 1.25 µm to 4.5 µm with a pulse duration in the femtosecond regime[27]. The light is transmitted through band-pass filters to select the desired part of the spectrum to be incident on the detector. An off-axis parabolic focusing mirror was used to direct the mid-infrared pulsed onto the device with spot sizes of 8 µm$^2$ and 11.5 µm$^2$ for the respective center wavelengths of 2.5 µm and 3.6 µm.

Figure 5 shows the impulse response measurements for a range of incident energy densities (10–200 µJ·cm$^{-2}$ for 1.56 µm and 2.5 µm center wavelengths, 2.5–30 µJ·cm$^{-2}$ for 3.6 µm center wavelength). The data show that the photoconductive effect persists to mid-infrared wavelengths as expected from the 0.3 eV bandgap in bulk black phosphorus. A time constant of order 1 ns is observed for all wavelengths measured and is consistent with the autocorrelation measurements carried out at 1.56 µm, indicating that the detector approaches the maximum bandwidth set by the intrinsic carrier recombination time scale.

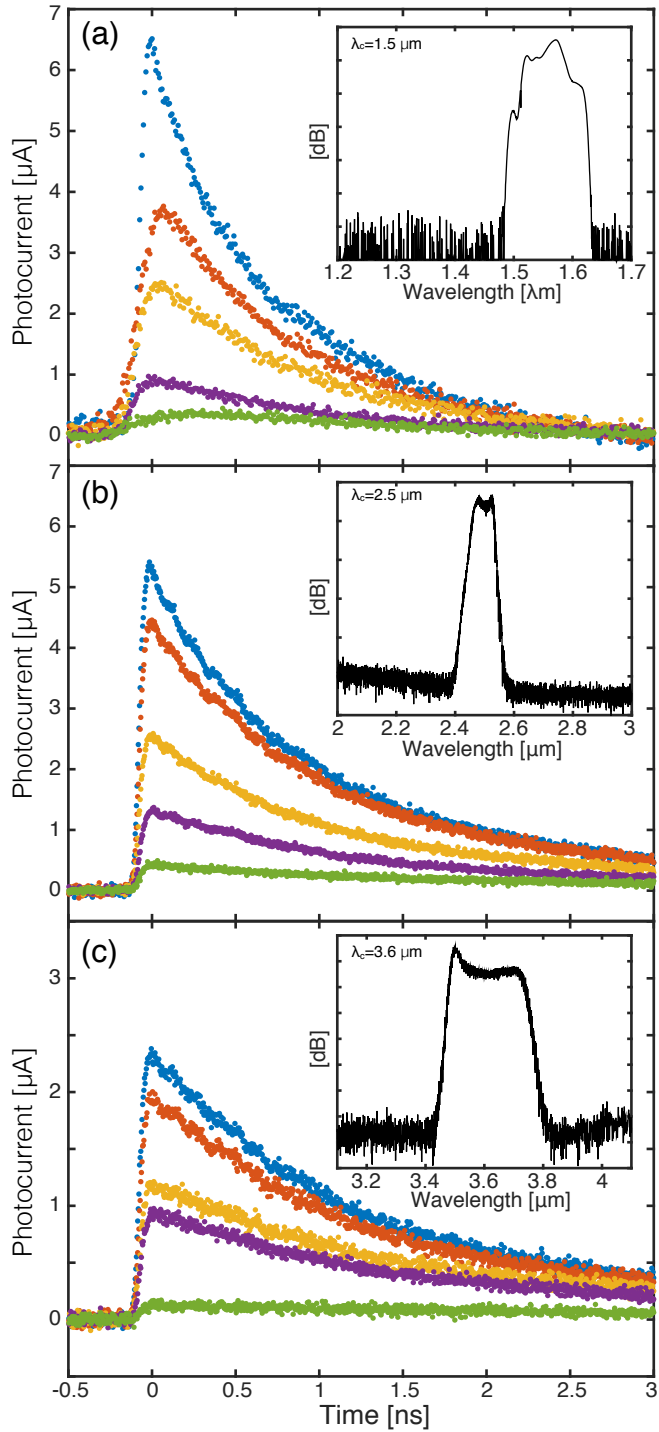

Figure 5. Impulse response data for black phosphorus detector showing photocurrent as a function of time for a range of incident powers (colors). The inset in (a-c) shows the corresponding spectrum which spans the near- to mid-infrared frequency range.

Rise times (10% to 90%) were estimated from the leading edge of the impulse response data shown in Fig. 5 and were found to be 65 ps for all measured wavelengths (cf., Fig. 6(a)) while Fig. 6(b) shows the peak photocurrent for the same data plotted as a function of total applied flux. The dashed line shows the photocurrent generated by the change in conductivity is initially linear with incident flux as expected for direct transition photoabsorption. Since the peak photocurrent is plotted against total incident flux, the vertical offset of the two dashed lines indicates differences in the intercepted flux between the different wavelengths, due to differences in spot size, as well as possible wavelength-dependent absorption.

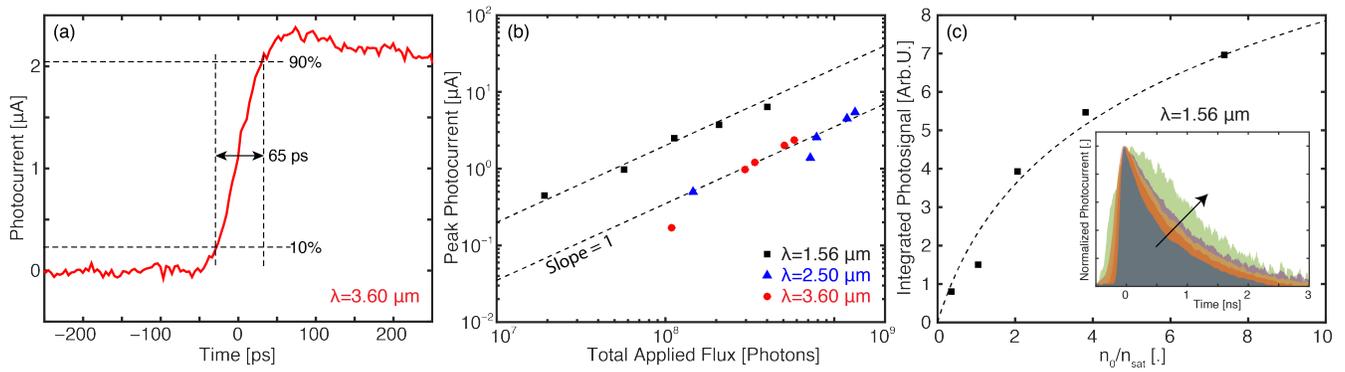

Figure 6. (a) Photocurrent impulse response showing 65 ps rise time (3.6 µm center wavelength shown). (b) Peak photocurrent for all wavelengths plotted as a function of incident flux. For all measurements a bias voltage of 200 mV and back gate of 0 V was used. (c) Average photocurrent in arbitrary units produced by integrating the near-infrared impulse response versus normalized excess carrier density (symbols). The dashed curve is the characteristic expected from the excess carrier model. The inset visualizes the photocurrent dependence on incident power by plotting the normalized impulse response taken at different illumination intensities (direction of arrow indicates decreasing fluence).

Figure 6(c) shows the average signal resulting from the integration of the impulse response data shown in Fig. 5(a). Even though the peak value increases linearly with increased photon flux as seen in Fig. 6(b), the average current measured is proportional to the total current that flows during the photoinduced conductivity transient. Figure 6(c) plots the area of the impulse response data in arbitrary units as a function of normalized excess carrier density. The data exhibits the same saturation characteristic as the single beam measurements presented in Fig. 3 and matches well to a fit line using the excess carrier model given by Eqn. 2.

**Discussion**

Many detectors currently available for the mid-infrared region employ low-gap semiconductors, such as HgCdTe [28] or PbSe [29]. These types of detectors typically require cooling below room temperature to reduce the inherently high dark conductivity caused by thermally activated carriers. In general, MIR detectors usually are optimized for a fast response, or a high responsivity, depending on the purpose of the device. Recently developed PbSe detectors allow room temperature operation at high detectivity values of $2.8 \times 10^{10}$ cm·Hz$^{1/2}$·W$^{-1}$, but have bandwidths limited to several kHz[30]. High speed MIR detectors based on HgCdTe allow for detection on the nanosecond scale, but suffer from a low detectivity[31]. Another class of detectors for the MIR wavelength range, based on superlattices[32] or quantum dots[33], can achieve high detectivity, but require cooling for operation. In contrast to graphene, thin layers of black phosphorus feature a low bandgap of 0.3 eV, making it naturally suited for efficient MIR detection for a range of wavelengths having photon energies above the bandgap. Electrostatic gating allows the conductivity of the flake to be minimized which suppresses the dark current in the device. The dark current for the data shown in Fig. 5 is 40 µA, though values of order 100 nA can be achieved if the gate voltage is tuned to the charge neutral point (cf., Fig. 1(b)) and modest bias voltages in the range of 10's of mV are used. Unlike conventional MIR detection materials, black phosphorous can be readily transferred to other passive or active optoelectronic devices or circuits, greatly increasing its versatility[34]. The noise equivalent power of 530 pW·Hz$^{-1/2}$ measured in our study corresponds to a detectivity of $10^6$ cm·Hz$^{1/2}$·W$^{-1}$ which is the expected value for Johnson noise limited performance. As the rise time of the black phosphorus detector is in the picosecond range, the limiting factor for the temporal resolution is set by the carrier life time which is of order 1 ns[25]. A high applied field across the channel and optimized flake orientation could further reduce the carrier lifetime by sweeping carriers out of the device[26], potentially allowing for sub-nanosecond carrier lifetimes.

**Conclusion**

We presented a detector that utilized the photoconductive effect in black phosphorus to detect mid-infrared light at room temperature with nanosecond-scale response times and thermally limited noise performance. Electrical transport and KPFM measurements confirm photoconductive operation, while autocorrelation measurements indicated the detection mechanism was sublinear with power, owing to a carrier recombination time that decreases with absorbed energy – an effect that is well described by a nonlinear excess carrier model. Impulse response measurements revealed a fast rise time of 65 ps and a carrier recombination time on the order of 1 ns from the near- to mid-infrared (1.56 μm to 3.75 μm). Using thinner flakes with optimum orientation to increase the maximum applied field tolerance and mobility could be used to further reduce the carrier lifetime and allow for bandwidths exceeding 1 gigahertz. Our results establish that the photoconductive effect in black phosphorus is an attractive platform for sensitive and fast mid-infrared detection.

**Acknowledgements**

Parts of this work were supported by the Office of Naval Research (ONR) Award No. N000141310865 and the National Science Foundation (NSF) Award No. ECCS1309750. The sample fabrication was carried out at the University of Maryland Nanocenter.